\documentclass{emulateapj}

\def\tobs{t_{\rm obs}}

\def\nuobs{\nu_{\rm obs}}

\def\be{\begin{equation}}
\def\ee{\end{equation}}
\def\beq{\begin{eqnarray}}
\def\eeq{\end{eqnarray}}

\begin{document}

\title{Evidence of Bulk Acceleration of the GRB X-ray Flare Emission Region}

\author{Z. Lucas Uhm, Bing Zhang}
\affil{Department of Physics and Astronomy, University of Nevada, Las Vegas, NV 89154, USA}

\begin{abstract}
Applying our recently-developed generalized version of the high-latitude emission theory to the observations of X-ray flares in gamma-ray bursts (GRBs), we present here clear observational evidence that the X-ray flare emission region is undergoing rapid bulk acceleration as the photons are emitted. We show that both the observed X-ray flare light curves and the photon index evolution curves can be simultaneously reproduced within a simple physical model invoking synchrotron radiation in an accelerating emission region far from the GRB central engine. Such an acceleration process demands an additional energy dissipation source other than kinetic energy, which points towards a significant Poynting-flux in the emission region of X-ray flares. As the X-ray flares are believed to share a similar physical mechanism as the GRB prompt emission, our finding here hints that the GRB prompt emission jets may also carry a significant Poynting-flux in their emitting region.
\end{abstract}


%
%

\section{Introduction}

Gamma-ray bursts (GRBs), the most luminous explosions in the universe, invoke relativistic jets beaming towards Earth with the highest velocities for bulk motion in the universe \citep[e.g.,][]{kumarzhang15}. Some of them are followed by softer, less energetic, X-ray flares, which also move with relativistic velocities towards Earth \citep{burrows05}. The X-ray flares are detected by the {\it Swift} satellite at $\sim (10^2-10^5)$ seconds after the GRB trigger, show a rapid rise and a steep fall, outshine the underlying afterglow by up to $\sim 3$ orders of magnitude in energy flux, and display a distinctive pattern of spectral evolution with a strong spectral softening during the decay phase. All these features are inconsistent with the standard afterglow emission, emitted from an external blast wave that propagates through a surrounding medium \citep{meszarosrees97,sari98}.
Indeed, observations \citep[e.g.,][]{liang06,chincarini07,margutti10} and theoretical modelings \citep[e.g.,][]{zhang06,lazzati07,maxham09} rather suggest that the X-ray flares share a similar physical mechanism as the GRB prompt emission itself, which results from internal energy dissipation within the jet.

For a spherical, relativistic jet, the decay of light curve cannot be steeper than a certain level defined by the so-called ``high-latitude emission effect'', or the ``curvature effect''. This is to say, even if the emission stops abruptly, photons from higher latitudes with respect to the observer's line of sight would arrive the detector at progressively later epochs but with a progressively lower Doppler factor 
value, resulting in a steep decay in flux. If the emission region keeps a constant Lorentz factor $\Gamma$, there exists a simple relation \citep{kumar00}
\begin{equation}
\label{eq:alpha_beta}
\hat \alpha = 2+\hat \beta
\end{equation}
between the temporal index $\hat \alpha$ and the spectral index $\hat \beta$, in the convention that the observed spectral flux is expressed as $F_{\nuobs}^{\, \rm obs} \propto \tobs^{-\hat \alpha}\, \nuobs^{-\hat \beta}$, where $\tobs$ is the observer time and $\nuobs$ is the observed frequency. Such a curvature effect has been invoked to interpret the fast decline of the X-ray flare light curves \citep{liang06}.

In this Letter, we confront the steep decay phase of X-ray flares with our generalized version of the high-latitude emission theory \citep{uhm15a} and show that the emitting region of X-ray flares undergoes rapid bulk acceleration. Also, for the first time, we present a physical modeling to both the observed flare light curves and the photon index evolution curves, simultaneously.

%
%

\section{High-latitude curvature effect and the steep decay phase of X-ray flares} \label{section:2}

In practice, testing the curvature effect theory with data is obscured by the so-called ``$T_0$-effect'' \citep{zhang06}. Since GRB light curves are plotted in logarithmic scale for both the observer time and the flux, the apparent decay slope $\hat\alpha_{\rm obs}$ sensitively depends on the reference time ($t_{\rm obs}=0$) to plot the light curves. Usually $t_{\rm obs}=0$ is defined at the GRB trigger time. For an X-ray flare, however, the emission episode likely starts at a later time (say, $t_{\rm obs} = T_0$) than prompt emission, so that a new zero time ($t_{\rm obs}=0$ at $T_0$) should be chosen in order to correctly study the radiation physics of the flare. Keeping the GRB trigger time as time zero point would cause an un-physically rapid decay that is steeper than the theoretical prediction. Indeed, the observed photon index $\hat \Gamma$ in the XRT band during the decay phase of an X-ray flare is typically in the range of $\sim (1 - 3)$, so that the corresponding spectral index, $\hat \beta = \hat \Gamma - 1$, is in the range $\sim (0 - 2)$. The expected temporal index $\hat \alpha$ from Equation (\ref{eq:alpha_beta}) should then be in the range $\sim (2 - 4)$, yet the observed decay indices $\hat\alpha_{\rm obs}$ are usually significantly larger than these values (up to more than 10). This has been attributed to the $T_0$ effect, suggesting that the X-ray flares have later emission episodes than the GRB prompt emission \citep{liang06}.

Correctly accounting for the $T_0$ effect is not straightforward, because the true beginning of the flares cannot be identified directly from the data. Depending on the detector sensitivity, the actual beginning may have been missed since it is too faint to be detected. More importantly, the initial portion of the flares is usually buried under the background afterglow emission. Thus, one may formally write that the beginning of an X-ray flare is at
\begin{equation}
\label{eq:T0}
T_0 = T_0^{\rm obs} - (\Delta T)^{\rm missed},
\end{equation}
where $T_0^{\rm obs}$ is the ``observed'' starting time of the flare and $(\Delta T)^{\rm missed}$ is the time duration of the missed portion. If one defines a new zero time $t_{\rm obs}=0$ at $T_0$, the new light curve of the flare would exhibit a shallower decay than the original observed one. The larger the $T_0$ value one corrects for, the shallower the decay slope one would get. The true $T_0$ value cannot be constrained observationally due to the unknown quantity $(\Delta T)^{\rm missed}$ but may be determined through theoretical modeling.

\begin{figure*} \centering
\centering
\begin{tabular}{ccc}
\includegraphics[width=5.7cm]{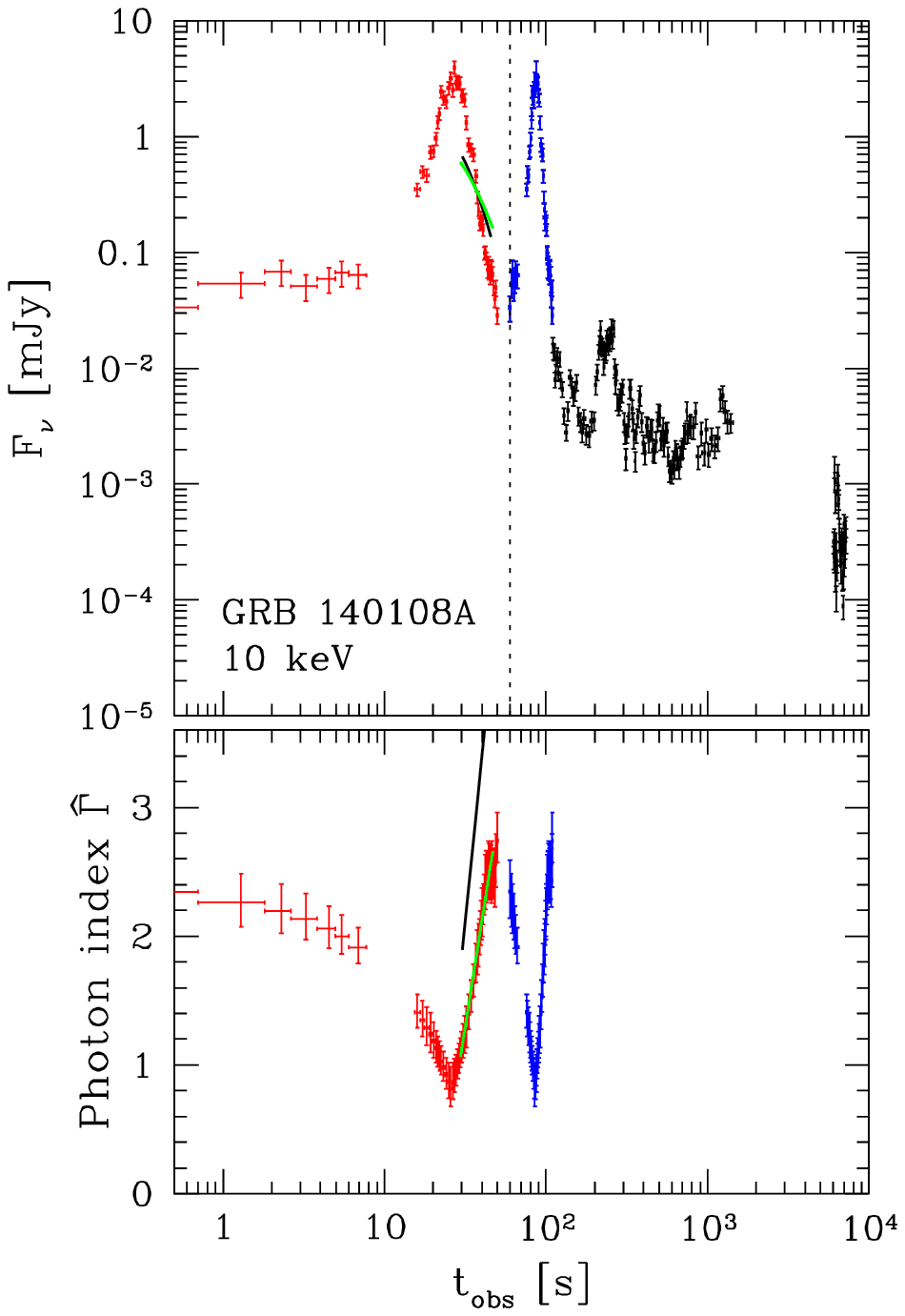} &
\includegraphics[width=5.7cm]{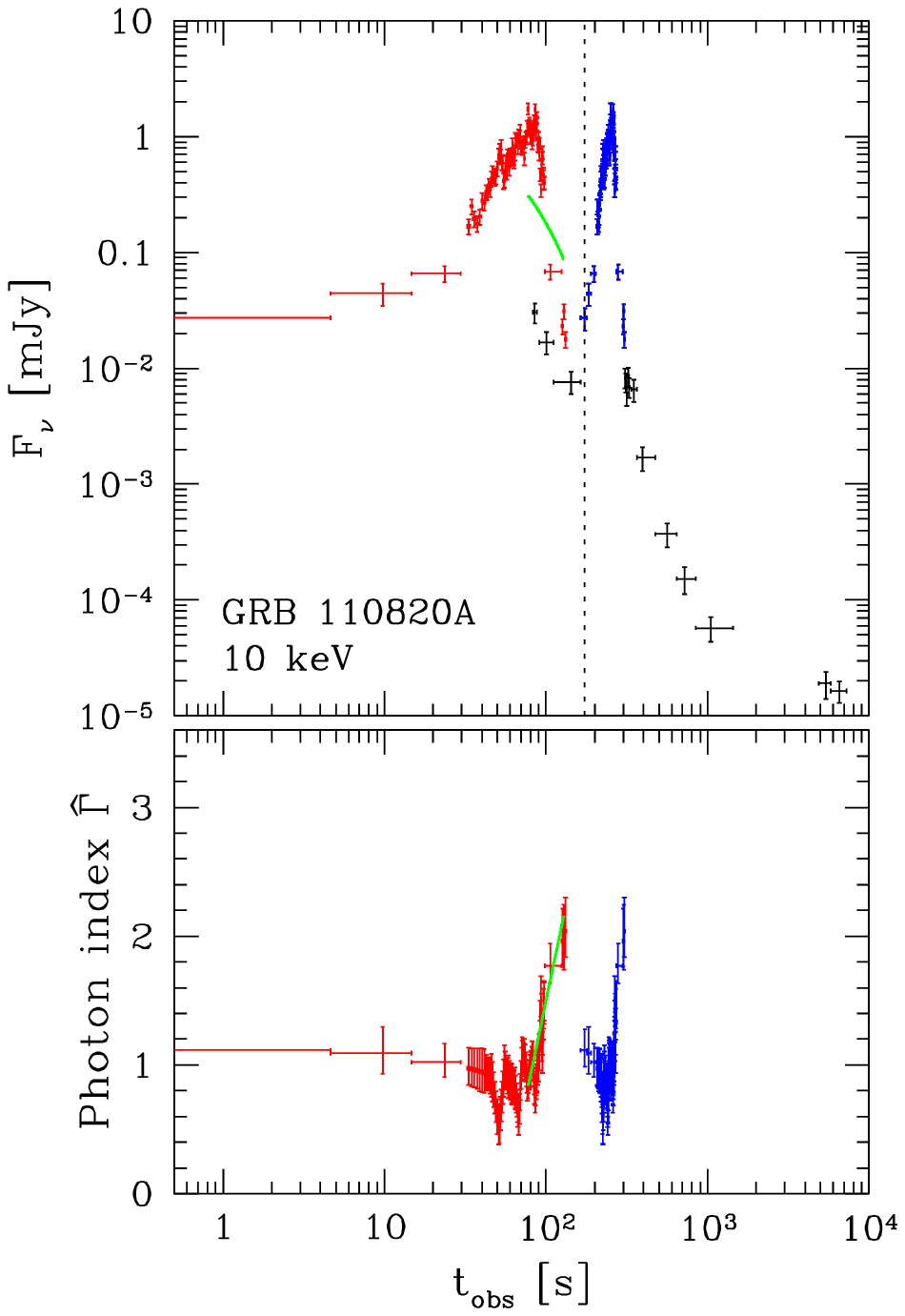} &
\includegraphics[width=5.7cm]{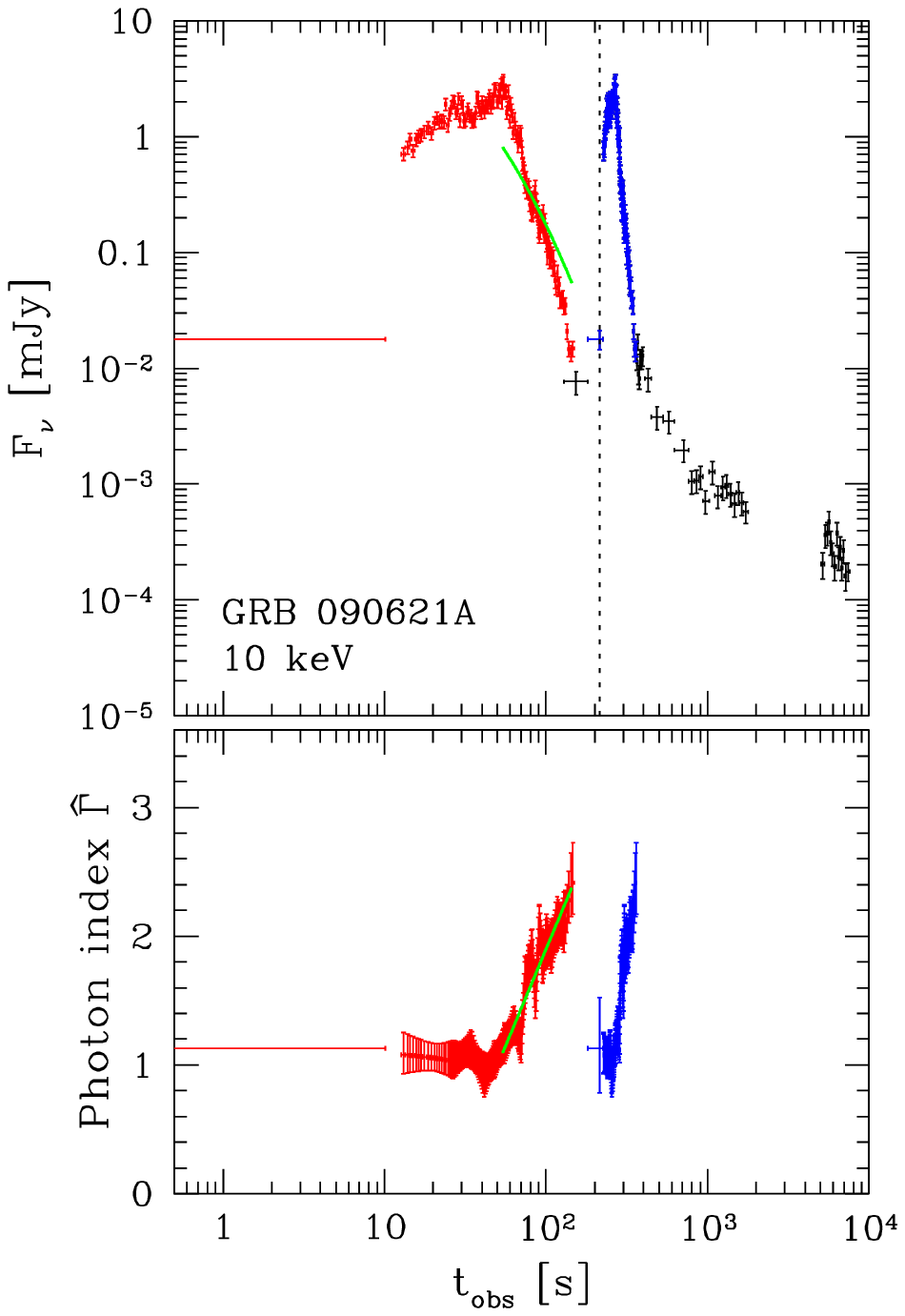} \\
\end{tabular}
\caption{
Three example X-ray flares in GRB 140108A, GRB 110820A, and GRB 090621A, that show strong observational evidence of significant bulk acceleration in the X-ray flare emission region. In the top panels, the black and blue data show the 10 keV spectral flux light curve as observed by XRT, with the blue color being used to identify the X-ray flare part. The blue data in the bottom panels show the photon index of the XRT band during the identified flare region. The dotted vertical line in the top panels represents the beginning of the observed data for each X-ray flare, i.e., $T_0^{\rm obs}$. We translate blue data points backwards by the amount $T_0^{\rm obs}$ and obtain red data points in top and bottom panels, respectively. By fitting the photon index evolution during the decay phase of flares (green curve in the bottom panels), we construct the predicted decay light curve (arbitrary normalization) based on Equation (\ref{eq:alpha_beta}) and show it by the green curve in the top panels. A more stringent approach may be possible through detailed modeling, as indicated by the black curves in GRB 140108A X-ray flare. See Section~\ref{section:2} for details.
All the original data are taken from the UK {\em Swift} Science Data Centre at the University of Leicester \citep{evans07}.
}
\label{fig:f1}
\end{figure*}

We analyze three X-ray flares observed in GRB 140108A, GRB 110820A, and GRB 090621A (see Figure~\ref{fig:f1}). The data are taken directly from the UK {\it Swift} Science Data Centre at the University of Leicester \citep{evans07}, and the 10 keV $F_\nu$ light curves (top panels) and XRT-band (0.3-10 keV) photon index curves (bottom panels) are presented. By identifying a flare (blue data points) from an original light curve of each example, we define $T_0^{\rm obs}$ as the first data point of the flare and mark it with the dotted vertical line in the top panel. We first take $T_0$ equal to $T_0^{\rm obs}$, which is the upper limit value of $T_0$. By systematically subtracting $T_0$ from $t_{\rm obs}$ of each data point of the flare, we effectively re-define the zero time point and ``translate backwards'' the data points to construct a new light curve of the X-ray flare (red data points). 
The observed photon index ($\hat\Gamma$) curve (blue data points) during the identified flare region and its corresponding backward translation (red data points) are also presented.

\begin{figure*} \centering
\centering
\begin{tabular}{ccc}
\includegraphics[width=5.7cm]{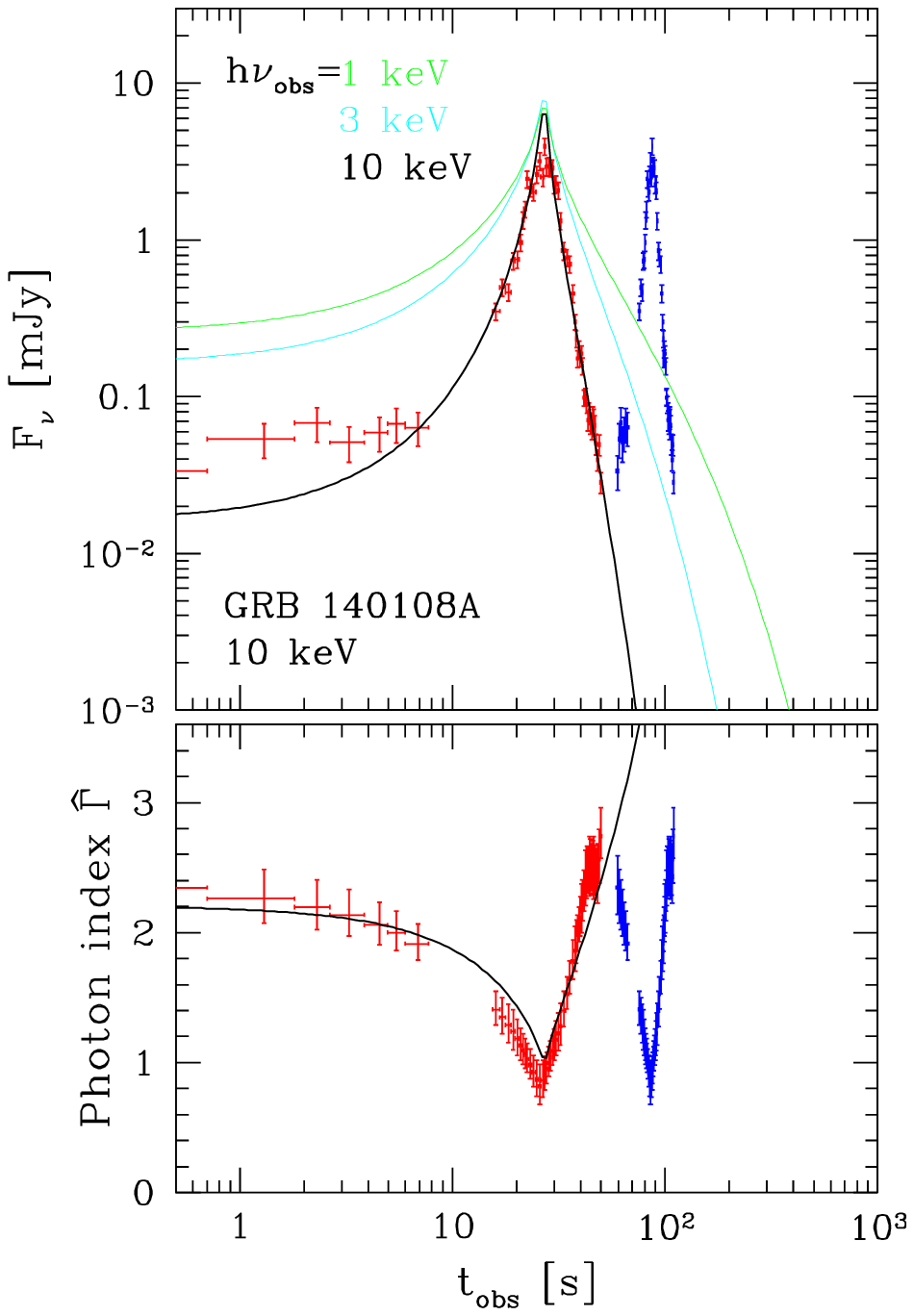} &
\includegraphics[width=5.7cm]{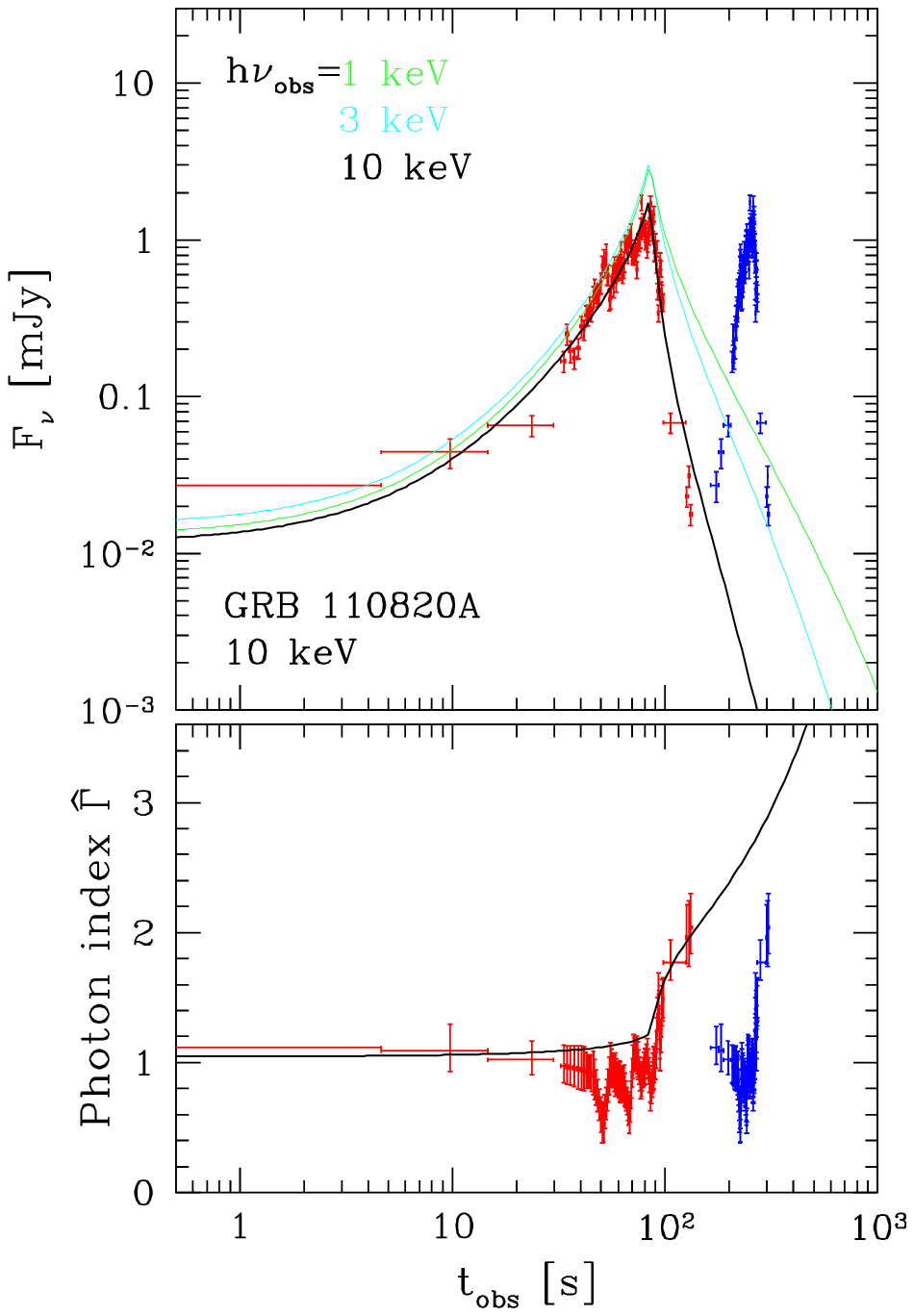} &
\includegraphics[width=5.7cm]{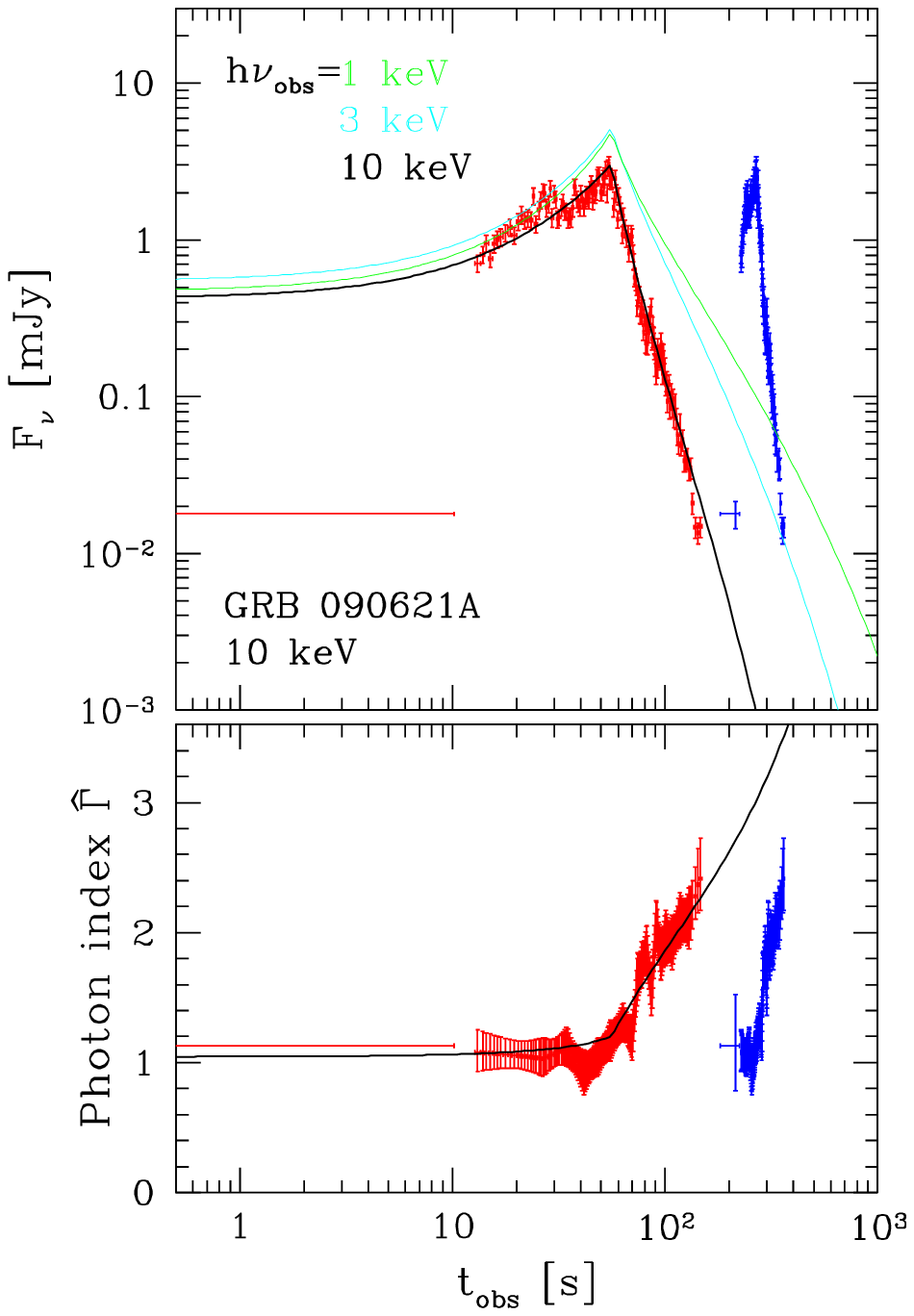} \\
\end{tabular}
\caption{
Modeling the three X-ray flares in GRB 140108A, GRB 110820A, and GRB 090621A, using the curvature effect with an accelerating emission region. Top panels present the observed flare light curve at 10 keV (red data points) and our model light curve at 10 keV (black curve). The model-predicted light curves at 1 keV (green) and 3 keV (cyan) are also presented. Bottom panels present a model prediction to the XRT-band photon index (black curve) and the observational data (red data points). See Section \ref{section:3} for details of modeling. The blue data points indicate the original observed light curve (top panel) and photon index curve (bottom panel) of each X-ray flare.
}
\label{fig:f2}
\end{figure*}

By fitting the photon index curve during the decay phase of the flare (green curve in the bottom panel), we reconstruct the predicted decay light curve (green curve in the top panel) based on Equation (\ref{eq:alpha_beta}), which has the steepest decay slope allowed by the curvature effect assuming that the jet is moving with a constant speed. One can see that the observed decay light curve violates this limit significantly in all three examples. 
Strictly speaking, $\hat\alpha$ should be derived by $\hat\beta$ at 10 keV (since the light curves are at 10 keV). The measured photon indices, and hence $\hat\beta$, on the other hand, are defined for the XRT band. A possible correction may be made through detailed modeling. Using the modeling results for the GRB 140108A X-ray flare (Figure~\ref{fig:f2} for details) as an example, we calculate the 10 keV time-dependent photon index (black curve in the bottom panel of GRB 140108A). Based on Equation (\ref{eq:alpha_beta}), the predicted decay light curve is constructed again (black curve in the top panel of GRB 140108A). Even though it is slightly steeper than the green curve, it is still significantly shallower than the observed light curve. Performing the same exercise to the other two X-ray flares leads to the same conclusion.

Notice that we have adopted the most conservative value for $T_0$ (i.e., its upper limit, $T_0^{\rm obs}$) in performing the test whether the observed decay phase of X-ray flares is consistent with Equation (\ref{eq:alpha_beta}).\footnote{In case where the photons are emitted from a region with a finite width, $T_0^{\rm obs}$ gives the upper limit value of $T_0$ only for the first shell in the emitting region while the test needs to be performed for the last shell. However, if an X-ray flare is produced from such a shell with a finite width, an expected feature of turning-off of the shell emission would be a rather flat shape of the peak area, due to the contributions from all shells in the region. All three examples in Figure~\ref{fig:f1} show a relatively sharp peak, without indicating such a flat feature at the peak area. Therefore, the analysis presented here remains valid.}
Correcting for a more realistic value (smaller than $T_0^{\rm obs}$) would lead to an even steeper decay curve after backward translation, worsening the conflict with Equation (\ref{eq:alpha_beta}) discussed above.

In a recent work on high-latitude emission theory \citep{uhm15a}, we find that Equation (\ref{eq:alpha_beta}) strictly holds only for a constant value of the Lorentz factor $\Gamma$. In the case of an accelerating shell, the temporal decay index $\hat \alpha$ becomes significantly larger than $2 + \hat \beta$. The trend is opposite for the case of a decelerating shell. Applying this new generalized version of the high-latitude emission theory to the apparent tension between the observational data and Equation (\ref{eq:alpha_beta}), we immediately conclude that {\em the emission regions of the three X-ray flares are undergoing significant bulk acceleration as the X-ray photons are emitted.}

%
%

\section{A simple physical model for X-ray flares} \label{section:3}

In order to verify the statement above, we perform detailed numerical modeling to these three X-ray flares shown in Figure~\ref{fig:f1}. We aim at reproducing the entire temporal behavior, i.e., not only the decaying phase but also the rising phase, for both the flare light curve and the photon index curve. The redshifts of the three GRBs are all unknown, and we assume a typical value $z=1$. The redshift plays only a global role in shaping the observed spectral flux \citep{uhm15a}, and therefore this assumption would not degrade the outcome of our modeling. The standard flat $\Lambda$CDM universe with the parameters $H_0=71$ km $\mbox{s}^{-1}$ $\mbox{Mpc}^{-1}$, $\Omega_{\rm m}=0.27$, and $\Omega_{\Lambda}=0.73$ is adopted in our calculations.

We adopt a simple physical picture:
a single relativistic spherical shell expands radially with a profile of the bulk Lorentz factor $\Gamma(r)$ as a function of radius $r$. The shell starts to emit photons at radius $r_{\rm on}$ (and at the lab-frame time $t_{\rm on}$), and finishes emitting at radius $r_{\rm off}$. During the emission phase, photons are continuously emitted from all locations in the shell, with an isotropic angular distribution of the emitted power in its co-moving fluid frame. Considering synchrotron radiation as the radiation mechanism (as suggested by recent theoretical modeling and data analysis \citep{uhm14,zhangbb16}), we delineate the shape of the photon spectrum in the fluid frame in the form of \citep{uhm15a} 
\be
\label{eq:H_ensemble}
H(x)
\quad
\mbox{with}
\quad
x=\nu^{\prime}/\nu_{\rm ch}^{\prime},
\ee
with the comoving-frame characteristic synchrotron frequency
\be
\nu_{\rm ch}^{\prime}=\frac{3}{16}\, \frac{q_{\rm e} B}{m_{\rm e} c}\, \gamma_{\rm ch}^2,
\ee
where $m_{\rm e}$ and $q_{\rm e}$ are the mass and charge of the electron, respectively, and $c$ is the speed of light. The magnetic field strength $B$ and the characteristic Lorentz factor $\gamma_{\rm ch}$ of the electrons 
are measured in the fluid frame, co-moving with the shell. The observed spectral flux, $F_{\nu_{\rm obs}}^{\,\rm obs}$, is calculated based on Doppler transformations from the comoving frame to the observer frame, with the curvature effect fully taken into account \citep{uhm15a}. The number of radiating electrons in the shell is assumed to increase at an injection rate $R_{\rm inj}$ (measured in the fluid frame) from an initial value $N=0$.

If one resets the reference time at $T_0$, the observer-frame time $t_{\rm obs}$ of the flare can be physically connected to the lab-frame time $t$ through
\beq
t_{\rm obs} 
&=& \frac{1}{c} \left[r_{\rm on} + c(t-t_{\rm on}) - r \cos \theta \right] (1+z) \nonumber \\
&=& \left[ \left(t-\frac{r}{c}\, \mu \right) - \left(t_{\rm on}-\frac{r_{\rm on}}{c}\right) \right] (1+z),
\eeq
taking into account the point that the initial photons of the flare are not emitted at $r=0$, but rather at $r_{\rm on}$ at $t_{\rm on}$. Here, $\mu \equiv \cos \theta$, and $\theta$ is the polar angle measuring the latitude of emission location with respect to the observer's line of sight. Notice that the observer time $\tobs$ is, in fact, independent of $t_{\rm on}$ since the time $t$ is calculated as $t=t_{\rm on}+\int_{r_{\rm on}} dr/(c\beta)$, where $\beta=(1-1/\Gamma^2)^{1/2}$.

However, in practice, the true $T_0$ value cannot be constrained directly from the data. Hence, we may instead reset the reference time at $T_0^{\rm obs}$, i.e., at the observed starting time of the flare. Then, according to Equation (\ref{eq:T0}), the observer time of the flare should read
\be
\label{eq:tobs_missed}
\tobs = \left[ \left(t-\frac{r}{c}\, \mu \right) - \left(t_{\rm on}-\frac{r_{\rm on}}{c}\right)\right] (1+z)-(\Delta T)^{\rm missed}.
\ee
This observer time is supposed to describe the red data points in Figure~\ref{fig:f1}, since those red points were obtained by reseting the reference time at $T_0^{\rm obs}$. Thus, we use Equation (\ref{eq:tobs_missed}) and model the red data points theoretically, which would enable us to constrain the unknown missed portion $(\Delta T)^{\rm missed}$.

In our modeling, we introduce power-law dependences of the following parameters:
\begin{eqnarray}
 \Gamma(r) &  = & \Gamma_0 \left(\frac{r}{r_0}\right)^s, \\
 B(r) & = & B_0 \left(\frac{r}{r_0}\right)^{-b}, \\
 \gamma_{\rm ch}(r) & = & \gamma_{\rm ch}^{0} \left(\frac{r}{r_0}\right)^g.
\end{eqnarray} 
The index $s$ delineates the degree of acceleration. The index $b$, with a typical value of 1, is naturally expected due to $B$-field flux conservation in an expanding shell, which is crucial to interpret GRB spectrum in the fast-cooling regime \citep{uhm14,zhangbb16}. The index $g$ describes how the characteristic electron Lorentz factor evolves with radius $r$. The shape of the photon spectrum $H(x)$ may be taken as a cutoff power law, or the so-called ``Band'' function \citep{band93}, which usually well describes the spectra of GRBs \citep{preece00} and can be accounted for within the fast-cooling synchrotron radiation model \citep{uhm14,zhangbb16}.

We begin our calculations at the radius $r_{\rm on}$. By setting $r_0 = r_{\rm on}$ and taking $b=1$ and $B_0 = 300$ G, we search for $s$, $g$, $r_{\rm off}$, $R_{\rm inj}$, $H(x)$, and $(\Delta T)^{\rm missed}$ to reproduce the observation of each of the three X-ray flares. The results are shown in Figure~\ref{fig:f2}. In all three examples, one can see a good agreement between our modeling and the data for both the 10keV-$F_\nu$ light curves (top panels) and the photon index evolution curves (bottom panels). The model-predicted light curves at 1 keV (green) and 3 keV (cyan) are also presented in the top panels. A time-dependent model-prediction curve for the XRT-band photon index is calculated as a power-law index between 1 keV and 10 keV light curves (black curve in the bottom panels). We note that without introducing the acceleration index $s$, the light curves and the photon index curves could not be reproduced. This suggests that our conclusion of an accelerating emission region for the X-ray flares in GRB 140108A, GRB 110820A, and GRB 090621A is robust.

Following parameters are adopted in common in all three examples: $r_{\rm on} = r_0 = 10^{14}$ cm, $B_0=300$ G, and $b=1$. The spectral function is taken as a power law with an exponential cutoff, i.e., $H(x) = x^{\alpha+1}\, e^{-x}$ with $\alpha = -0.7$. A Band-function \citep{band93} input spectrum with a steep high-energy index gives a similar overall match to the data. Other model parameters for each example are as follows. 
For (GRB 140108A, GRB 110820A, GRB 090621A): $\Gamma_0=(7.5, 6.0, 7.0)$, $s=(1.15, 0.95, 0.8)$, $\gamma_{\rm ch}^{0}=(0.82, 2.2, 2.2) \times 10^4$, $g=(0.48, 0.0, 0.05)$, $r_{\rm off}=(2.0, 2.0, 1.5) \times 10^{15}$ cm, and a constant injection rate, $R_{\rm inj}=(6.0, 2.5, 5.5) \times 10^{47} ~ \mbox{s}^{-1}$. A non-zero $(\Delta T)^{\rm missed}$ is indeed required to model the data, and for each case, $(\Delta T)^{\rm missed}=(18, 12, 36)$ s is adopted.

The rising phase of the flare in GRB 140108A exhibits a strong spectral hardening; the photon index decreases from $\hat \Gamma \simeq 2.4$ to $\hat \Gamma \simeq 1$.
This implies that the peak energy of the observed photon spectrum rapidly increases during this phase. According to our modeling, the spherical shell is actively emitting during the rising phase, and the temporal behavior of the observed peak energy roughly follows $E_{\rm p}^{\rm obs} \sim \Gamma B \gamma_{\rm ch}^2 \sim r^{s-b+2g}$. For the flare in GRB 140108A, the result of our modeling gives $E_{\rm p}^{\rm obs} \sim r^{1.11}$, confirming the rapid evolution of the peak energy during the rising phase. On the other hand, the strong spectral softening observed during the decay phase of all three flares is due to the high-latitude curvature effect in the acceleration regime \citep{uhm15a}. The emission from higher latitudes has a progressively smaller Doppler boosting, and therefore the observed spectrum is placed with a progressively smaller peak energy, resulting in a spectral softening.

%
%

\section{Conclusions and Discussion}

In this Letter, by applying our newly-developed generalized version of the high-latitude emission theory \citep{uhm15a} to the observations of three example X-ray flares in GRB 140108A, GRB 110820A, and GRB 090621A, we have presented clear observational evidence that the emitting region of these X-ray flares should undergo rapid bulk acceleration as the flare photons are emitted. Furthermore, we have shown that the entire observed temporal behavior, i.e., not only the decaying phase but also the rising phase, for both the flare light curves and the photon index evolution curves can be simultaneously reproduced within a simple physical model invoking synchrotron radiation in an accelerating emission region far ($\sim 10^{15}$ cm) from the GRB central engine. This is the first time that such a comprehensive theoretical modeling is done for GRB X-ray flares.

The identification of an acceleration process in a relativistic jet has profound implications. One important opening question in the field of GRBs is the composition of the jets \citep{kumarzhang15}. Within the standard matter-dominated ``fireball'' model, the jet undergoes rapid acceleration early on \citep{meszaros93,piran93,kobayashi99} below the coasting radius $r_c < \Gamma R_0 = 3\times 10^{11}~{\rm cm} (\Gamma/300) (R_0/10^9~{\rm cm})$ (where $R_0$ is the radius of the jet base), and becomes kinetic energy dominated afterwards. The non-thermal emission is believed to be emitted from internal shocks where the kinetic energy is dissipated \citep{rees94}. Within this scenario, the emission is released at the expense of the kinetic energy, so that after the prompt emission phase, the average Lorentz factor of the fireball is expected to be reduced. An alternative scenario invokes a Poynting-flux-dominated outflow, with the magnetization parameter $\sigma$ (the ratio between Poynting flux and matter flux) greater than unity in the emission region. It has been known that such kind of jet may undergo slow acceleration even without magnetic dissipation and photon radiation \citep{komissarov09,granot11}, as the Poynting flux energy is gradually converted to kinetic energy due to the magnetic pressure gradient within the jet. If the Poynting flux undergoes an abrupt dissipation, probably due to internal collision-induced magnetic reconnection and turbulence (ICMART) \citep{zhangyan11,deng15}, part of the dissipated Poynting flux energy would be given to the jet for bulk acceleration as the other portion of the energy is converted to particle energy and released as photons. Such a model also invokes a relatively large emission radius from the central engine \citep{zhangyan11}, which is required to interpret the long-duration X-ray flare decay tail within the curvature effect model \citep{zhang06}. As a result, our finding provides a ``smoking-gun'' signature of $\sigma > 1$ in the emission region of X-ray flares. 
This aligns with other arguments that X-ray flares are Poynting-flux-dominated \citep{fan05e}.

Observational and theoretical arguments suggest that GRB prompt emission shares a similar physical origin with the X-ray flare emission \citep{liang06,chincarini07,margutti10,zhang06,lazzati07,maxham09}. 
Therefore, our finding of bulk acceleration in X-ray flares hints that the emission region of GRBs may also have $\sigma > 1$, which is consistent with other observational evidence, including the lack of or the weak thermal emission component in most GRBs \citep{zhangpeer09,gaozhang15}, the polarized $\gamma$-ray emission \citep{yonetoku11}, and the polarized early optical afterglow emission from the reverse shock\footnote{We expect that most of the Poynting-flux energy would be largely dissipated or consumed while producing the bright prompt gamma-rays or X-ray flares. Therefore, when such a consumed outflow enters an afterglow stage, the reverse-shock emission is not necessarily suppressed. Rather, one may expect an enhanced reverse-shock emission or a polarized optical emission from the reverse shock due to the remaining weak magnetic fields in the outflow.} \citep{steele09,mundell13}.
Indeed, based on an independent argument regarding the observed spectral lags, we recently showed that the GRB prompt-emission region is also undergoing bulk acceleration \citep{uhm16a}.

A decay index $\hat\alpha$ steeper than $2+\hat\beta$ can also be achieved by invoking anisotropic emission in the jet co-moving frame \citep{beloborodov11b}. However, both spectral and temporal properties of X-ray flares, i.e., strong spectral evolution observed during both the rising and decaying phases of flares as well as steep decay light curves, cannot be easily interpreted within such a scenario. Bulk acceleration is demanded in order to simultaneously interpret the light curve and spectral evolution of the flares.
An extended analysis \citep{jia15} shows that the majority of X-ray flares are in the acceleration regime, suggesting a ubiquitous Poynting-flux-dominated composition among X-ray flares.

%
%

\acknowledgments
We thank Pawan Kumar, Yizhong Fan, Judith Racusin, and the anonymous referee for helpful comments and discussion. 
This work is supported by NASA through an Astrophysical Theory Program (grant number NNX 15AK85G) and an Astrophysics Data Analysis Program (grant number NNX 14AF85G). It made use of data supplied by the UK Swift Science Data Centre at the University of Leicester.

%
%


\begin{thebibliography}{34}
\expandafter\ifx\csname natexlab\endcsname\relax\def\natexlab#1{#1}\fi

\bibitem[{{Band} {et~al.}(1993){Band}, {Matteson}, {Ford}, {Schaefer},
  {Palmer}, {Teegarden}, {Cline}, {Briggs}, {Paciesas}, {Pendleton}, {Fishman},
  {Kouveliotou}, {Meegan}, {Wilson}, \& {Lestrade}}]{band93}
{Band}, D., {Matteson}, J., {Ford}, L., {et~al.} 1993, \apj, 413, 281

\bibitem[{{Beloborodov} {et~al.}(2011){Beloborodov}, {Daigne}, {Mochkovitch},
  \& {Uhm}}]{beloborodov11b}
{Beloborodov}, A.~M., {Daigne}, F., {Mochkovitch}, R., \& {Uhm}, Z.~L. 2011,
  \mnras, 410, 2422

\bibitem[{{Burrows} {et~al.}(2005){Burrows}, {Romano}, {Falcone}, {Kobayashi},
  {Zhang}, {Moretti}, {O'Brien}, {Goad}, {Campana}, {Page}, {Angelini},
  {Barthelmy}, {Beardmore}, {Capalbi}, {Chincarini}, {Cummings}, {Cusumano},
  {Fox}, {Giommi}, {Hill}, {Kennea}, {Krimm}, {Mangano}, {Marshall},
  {M{\'e}sz{\'a}ros}, {Morris}, {Nousek}, {Osborne}, {Pagani}, {Perri},
  {Tagliaferri}, {Wells}, {Woosley}, \& {Gehrels}}]{burrows05}
{Burrows}, D.~N., {Romano}, P., {Falcone}, A., {et~al.} 2005, Science, 309,
  1833

\bibitem[{{Chincarini} {et~al.}(2007){Chincarini}, {Moretti}, {Romano},
  {Falcone}, {Morris}, {Racusin}, {Campana}, {Covino}, {Guidorzi},
  {Tagliaferri}, {Burrows}, {Pagani}, {Stroh}, {Grupe}, {Capalbi}, {Cusumano},
  {Gehrels}, {Giommi}, {La Parola}, {Mangano}, {Mineo}, {Nousek}, {O'Brien},
  {Page}, {Perri}, {Troja}, {Willingale}, \& {Zhang}}]{chincarini07}
{Chincarini}, G., {Moretti}, A., {Romano}, P., {et~al.} 2007, \apj, 671, 1903

\bibitem[{{Deng} {et~al.}(2015){Deng}, {Li}, {Zhang}, \& {Li}}]{deng15}
{Deng}, W., {Li}, H., {Zhang}, B., \& {Li}, S. 2015, \apj, 805, 163

\bibitem[{{Evans} {et~al.}(2007){Evans}, {Beardmore}, {Page}, {Tyler},
  {Osborne}, {Goad}, {O'Brien}, {Vetere}, {Racusin}, {Morris}, {Burrows},
  {Capalbi}, {Perri}, {Gehrels}, \& {Romano}}]{evans07}
{Evans}, P.~A., {Beardmore}, A.~P., {Page}, K.~L., {et~al.} 2007, \aap, 469,
  379

\bibitem[{{Fan} {et~al.}(2005){Fan}, {Zhang}, \& {Proga}}]{fan05e}
{Fan}, Y.~Z., {Zhang}, B., \& {Proga}, D. 2005, \apjl, 635, L129

\bibitem[{{Gao} \& {Zhang}(2015)}]{gaozhang15}
{Gao}, H., \& {Zhang}, B. 2015, \apj, 801, 103

\bibitem[{{Granot} {et~al.}(2011){Granot}, {Komissarov}, \&
  {Spitkovsky}}]{granot11}
{Granot}, J., {Komissarov}, S.~S., \& {Spitkovsky}, A. 2011, \mnras, 411, 1323

\bibitem[{{Jia} {et~al.}(2015){Jia}, {Uhm}, \& {Zhang}}]{jia15}
{Jia}, L.-W., {Uhm}, Z.~L., \& {Zhang}, B. 2015, \apjs, in press
  (arXiv:1509.04871)

\bibitem[{{Kobayashi} {et~al.}(1999){Kobayashi}, {Piran}, \&
  {Sari}}]{kobayashi99}
{Kobayashi}, S., {Piran}, T., \& {Sari}, R. 1999, \apj, 513, 669

\bibitem[{{Komissarov} {et~al.}(2009){Komissarov}, {Vlahakis}, {K{\"o}nigl}, \&
  {Barkov}}]{komissarov09}
{Komissarov}, S.~S., {Vlahakis}, N., {K{\"o}nigl}, A., \& {Barkov}, M.~V. 2009,
  \mnras, 394, 1182

\bibitem[{{Kumar} \& {Panaitescu}(2000)}]{kumar00}
{Kumar}, P., \& {Panaitescu}, A. 2000, \apjl, 541, L51

\bibitem[{{Kumar} \& {Zhang}(2015)}]{kumarzhang15}
{Kumar}, P., \& {Zhang}, B. 2015, \physrep, 561, 1

\bibitem[{{Lazzati} \& {Perna}(2007)}]{lazzati07}
{Lazzati}, D., \& {Perna}, R. 2007, \mnras, 375, L46

\bibitem[{{Liang} {et~al.}(2006){Liang}, {Zhang}, {O'Brien}, {Willingale},
  {Angelini}, {Burrows}, {Campana}, {Chincarini}, {Falcone}, {Gehrels}, {Goad},
  {Grupe}, {Kobayashi}, {M{\'e}sz{\'a}ros}, {Nousek}, {Osborne}, {Page}, \&
  {Tagliaferri}}]{liang06}
{Liang}, E.~W., {Zhang}, B., {O'Brien}, P.~T., {et~al.} 2006, \apj, 646, 351

\bibitem[{{Margutti} {et~al.}(2010){Margutti}, {Guidorzi}, {Chincarini},
  {Bernardini}, {Genet}, {Mao}, \& {Pasotti}}]{margutti10}
{Margutti}, R., {Guidorzi}, C., {Chincarini}, G., {et~al.} 2010, \mnras, 406,
  2149

\bibitem[{{Maxham} \& {Zhang}(2009)}]{maxham09}
{Maxham}, A., \& {Zhang}, B. 2009, \apj, 707, 1623

\bibitem[{{M\'esz\'aros} {et~al.}(1993){M\'esz\'aros}, {Laguna}, \&
  {Rees}}]{meszaros93}
{M\'esz\'aros}, P., {Laguna}, P., \& {Rees}, M.~J. 1993, \apj, 415, 181

\bibitem[{{M\'esz\'aros} \& {Rees}(1997)}]{meszarosrees97}
{M\'esz\'aros}, P., \& {Rees}, M.~J. 1997, \apj, 476, 232

\bibitem[{{Mundell} {et~al.}(2013){Mundell}, {Kopa{\v c}}, {Arnold}, {Steele},
  {Gomboc}, {Kobayashi}, {Harrison}, {Smith}, {Guidorzi}, {Virgili},
  {Melandri}, \& {Japelj}}]{mundell13}
{Mundell}, C.~G., {Kopa{\v c}}, D., {Arnold}, D.~M., {et~al.} 2013, \nat, 504,
  119

\bibitem[{{Piran} {et~al.}(1993){Piran}, {Shemi}, \& {Narayan}}]{piran93}
{Piran}, T., {Shemi}, A., \& {Narayan}, R. 1993, \mnras, 263, 861

\bibitem[{{Preece} {et~al.}(2000){Preece}, {Briggs}, {Mallozzi}, {Pendleton},
  {Paciesas}, \& {Band}}]{preece00}
{Preece}, R.~D., {Briggs}, M.~S., {Mallozzi}, R.~S., {et~al.} 2000, \apjs, 126,
  19

\bibitem[{{Rees} \& {M\'esz\'aros}(1994)}]{rees94}
{Rees}, M.~J., \& {M\'esz\'aros}, P. 1994, \apjl, 430, L93

\bibitem[{{Sari} {et~al.}(1998){Sari}, {Piran}, \& {Narayan}}]{sari98}
{Sari}, R., {Piran}, T., \& {Narayan}, R. 1998, \apjl, 497, L17+

\bibitem[{{Steele} {et~al.}(2009){Steele}, {Mundell}, {Smith}, {Kobayashi}, \&
  {Guidorzi}}]{steele09}
{Steele}, I.~A., {Mundell}, C.~G., {Smith}, R.~J., {Kobayashi}, S., \&
  {Guidorzi}, C. 2009, \nat, 462, 767

\bibitem[{{Uhm} \& {Zhang}(2014)}]{uhm14}
{Uhm}, Z.~L., \& {Zhang}, B. 2014, Nature Physics, 10, 351

\bibitem[{{Uhm} \& {Zhang}(2015)}]{uhm15a}
---. 2015, \apj, 808, 33

\bibitem[{{Uhm} \& {Zhang}(2016)}]{uhm16a}
---. 2016, \apj, in press (arXiv:1511.08807)

\bibitem[{{Yonetoku} {et~al.}(2011){Yonetoku}, {Murakami}, {Gunji}, {Mihara},
  {Toma}, {Sakashita}, {Morihara}, {Takahashi}, {Toukairin}, {Fujimoto},
  {Kodama}, {Kubo}, \& {IKAROS Demonstration Team}}]{yonetoku11}
{Yonetoku}, D., {Murakami}, T., {Gunji}, S., {et~al.} 2011, \apjl, 743, L30

\bibitem[{{Zhang} {et~al.}(2006){Zhang}, {Fan}, {Dyks}, {Kobayashi},
  {M{\'e}sz{\'a}ros}, {Burrows}, {Nousek}, \& {Gehrels}}]{zhang06}
{Zhang}, B., {Fan}, Y.~Z., {Dyks}, J., {et~al.} 2006, \apj, 642, 354

\bibitem[{{Zhang} \& {Pe'er}(2009)}]{zhangpeer09}
{Zhang}, B., \& {Pe'er}, A. 2009, \apjl, 700, L65

\bibitem[{{Zhang} \& {Yan}(2011)}]{zhangyan11}
{Zhang}, B., \& {Yan}, H. 2011, \apj, 726, 90

\bibitem[{{Zhang} {et~al.}(2016){Zhang}, {Uhm}, {Connaughton}, {Briggs}, \&
  {Zhang}}]{zhangbb16}
{Zhang}, B.-B., {Uhm}, Z.~L., {Connaughton}, V., {Briggs}, M.~S., \& {Zhang},
  B. 2016, \apj, 816, 72

\end{thebibliography}

%
%

\newpage

\end{document}